\documentclass{article}

    \PassOptionsToPackage{numbers, compress, sort}{natbib}

    \usepackage[final]{neurips_2024}

\usepackage[utf8]{inputenc} 
\usepackage[T1]{fontenc}    
\usepackage{hyperref}       
\usepackage{url}            
\usepackage{booktabs}       
\usepackage{amsfonts}       
\usepackage{nicefrac}       
\usepackage{microtype}      
\usepackage{xcolor}         
\usepackage{amsmath,graphicx}
\usepackage{caption}
\usepackage{subcaption}

\title{Challenge on Sound Scene Synthesis: \\Evaluating Text-to-Audio Generation}

\author{%
    Junwon Lee\thanks{Equal contribution} $^1$, Modan Tailleur$^{*2}$, Laurie M. Heller$^{*3}$, Keunwoo Choi$^{*4}$,\\
     \textbf{ Mathieu Lagrange$^{*2}$, Brian McFee$^5$, Keisuke Imoto$^6$, Yuki Okamoto$^7$}\\
    $^1$KAIST, $^2$Nantes Université,  $^3$CMU, $^4$Gaudio Lab, $^5$NYU, $^6$Doshisha Univ., $^7$UTokyo
}

\begin{document}

\maketitle

\begin{abstract}
Despite significant advancements in neural text-to-audio generation, challenges persist in controllability and evaluation. 
This paper addresses these issues through the \textit{Sound Scene Synthesis} challenge held as part of the Detection and Classification of Acoustic Scenes and Events 2024.  
We present an evaluation protocol combining objective metric, namely Fréchet Audio Distance,  with perceptual assessments, utilizing a structured prompt format to enable diverse captions and effective evaluation. Our analysis reveals varying performance across sound categories and model architectures, with larger models generally excelling but innovative lightweight approaches also showing promise. The strong correlation between objective metrics and human ratings validates our evaluation approach. We discuss outcomes in terms of audio quality, controllability, and architectural considerations for text-to-audio synthesizers, providing direction for future research.
\end{abstract}

\begin{figure}[h!]
    \centering
    \resizebox{0.8\textwidth}{!}{%
        \includegraphics{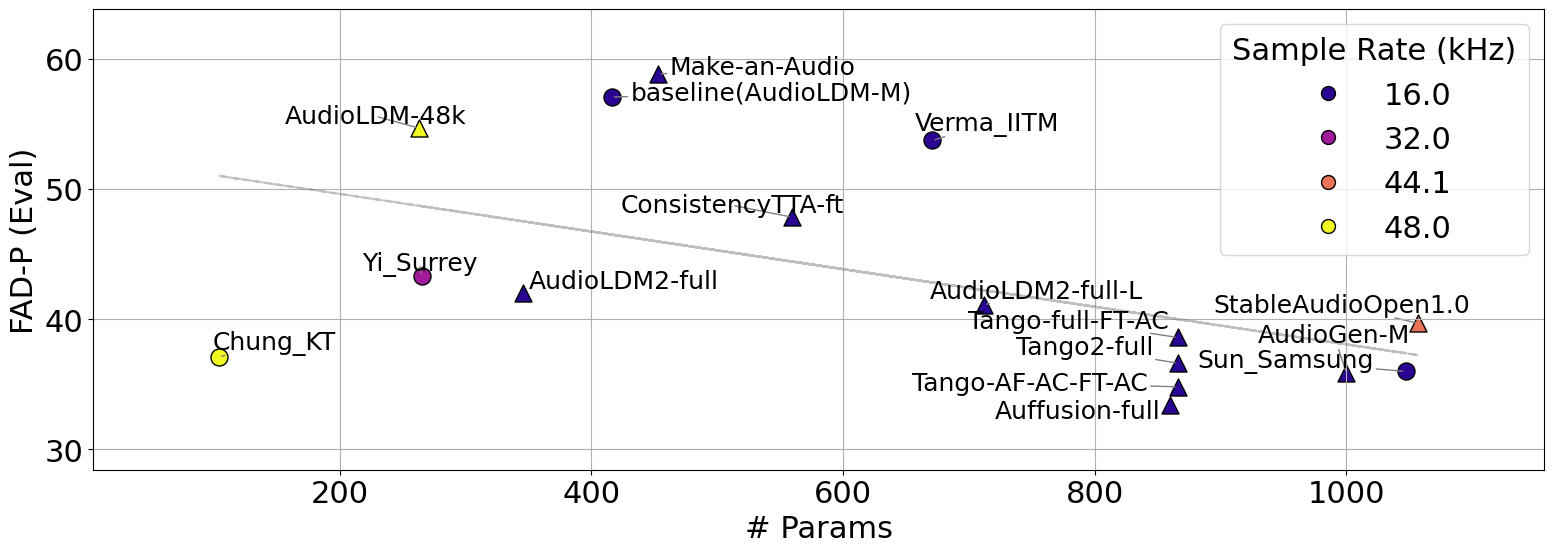}
    }
    \caption{Performance of various Text-to-Audio models (circled markers: challenge submissions, triangular markers: open-source models) on the evaluation set versus their number of parameters. Color depicts the audio sample rate. }
    \label{fig:fad-models}
\end{figure}

\section{Introduction}
\label{sec:introduction}

Sound is of paramount importance in the creation of an immersive user experience in multimedia content such as movies and games, not to mention real-time applications such as the metaverse. By generating sound that aligns with a target sound description, audio generation systems would offer creators a greater range of options and streamline the workflow, reducing time and cost. 

Recent advances in audio generation models have demonstrated considerable potential for automating and streamlining the process. 
However, the models face two significant challenges: a lack of audio quality that meets professional standards and a limited control over shaping desired sound sources.

To highlight such limitations and facilitate further research, we organized a challenge on sound scene synthesis. 
The protocol we proposed and followed includes guidelines for dataset construction, objective metrics, and a human evaluation scheme to answer these research questions:
1) How can a model generate high-fidelity(quality) audio
2) How can we improve the diversity of the generated sounds with diverse foreground and background sound sources
3) How can we enhance the controllability of the model to generate audio given a corresponding text caption
4) How can we evaluate the category appropriateness, perceptual quality, and diversity of model-generated sounds.

\section{Problem and Task Definition}
\label{sec:problem_and_task}
\begin{figure}[t!]
    \centering
    \resizebox{0.8\textwidth}{!}{%
        \includegraphics{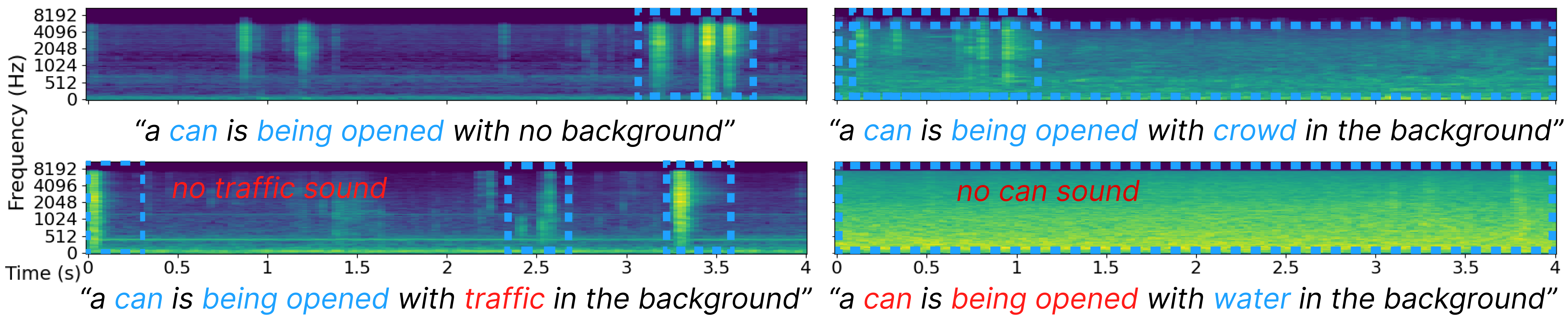}
    }
    \caption{Examples that show limited controllability of a recent text-to-audio model (AudioLDM-M \cite{audioldm}) while controlling sound sources.}
    \label{fig:text_prompt}
\end{figure}

In multimedia sound production, sound artists and engineers typically adhere to a structured process to create a final soundtrack. They first generate Foley sounds or collect samples from databases for each sound source. These samples are then edited to meet specific expectations regarding timbre, nuance, and temporal alignment. Finally, they mix all elements into a cohesive sound scene, often with music. TTA systems are designed to automate this process, but they face a number of challenges.

\subsection{Problem in Current Text-To-Audio Systems}
\label{sec:prob-in-current-tta}

First, the quality of the generated audio is usually inadequate to meet commercial standards. Many TTA systems \cite{audioldm,tango,audiogen,make-an-audio} generate audio waveforms at a 16 kHz sampling rate for training and inference efficiency, which is significantly lower than the industry standard of 48 kHz or higher. Second, their controllability through the text prompt is limited. Since controllability is crucial to achieve the desired sound characteristics, this limitation is a significant concern. Figure \ref{fig:text_prompt} illustrates how the well-known open-source TTA model (AudioLDM \cite{audioldm}) struggles with text-based controls. 
The generated sound is rarely aligned with both the foreground and background sounds, i.e., their \textit{compositionality} is noticeably limited. This happens particularly when there is a strong positive or negative correlation between the foreground and the background in the training set. 

Evaluation is another significant challenge, particularly because captions are incomplete descriptions of audio signals at varying levels of abstraction~\cite{audiocaps, wavcaps, clotho}. Fairly evaluating audio generation with a satisfaction score from such varied captioning styles presents considerable difficulties because of the seemingly endless possibilities of factors to consider. When evaluating a generated audio based on a caption \textit{People in a small crowd are speaking and a dog barks} (from AudioCaps), for example, should the number of people speaking be considered? 
Does "and" imply the sounds to be sequential or simultaneous? How should all these factors be weighted to compute the satisfaction score? Once we answer these questions, how can we aggregate the score of this example with a score of a much simpler prompt? 
Although it may not be practically possible to answer all these questions, a simplified protocol should be defined to organize a public challenge in a fair manner.

\subsection{Task Definition}
\label{sec:task-definition}
In general, sound scene synthesis refers to the task of generating environmental sound scenes that can accompany events in multimedia content to enhance the narrative experience, excluding speech and music. This  Sound Scene Synthesis task is built on last year’s Foley sound synthesis challenge 
\cite{dcase2023, choi2022proposalfoleysoundsynthesis}, 
expanding the scope from Foley sounds to general sound scenes by generalizing the conditioning from a single predefined sound category to a natural language prompt. The audio output requirement is a 4-second, 32-bit, 32kHz, mono-channel audio waveform. Each submitted model is required to generate 250 audio files within a 24-hour period using the computing environment of \textit{Colab Pro+}.

The evaluation prompts are limited to the following structure: \textit{'(foreground sound source) with (background sound source) in the background'}, with action-based foreground sounds and ambient background sounds specified within the parenthesis. This format was devised to enable quantified evaluation of diverse text prompts.

\section{Official Dataset and Baseline System}
\label{sec:dataset_and_baseline}

\textbf{Dataset Creation} Prompts following the structure described in section \ref{sec:task-definition} were crafted manually by the organizing team. We categorized foreground prompts into six categories: "animal," "vehicle," "human," "alarm," "tool," and "entrance." These foreground prompts are paired with five different backgrounds: "crowd," "traffic," "water," "birds," and "no background," except that vehicles are not paired with traffic. The "no background" permits evaluation of clean monophonic foreground audios.

The level of detail in prompts was adjusted depending on the nature of the sound source. For example, the foreground prompt "a jackhammer is pounding" provides a clear and self-sufficient description. Qualifiers such as "small" or "large" would contribute little to the perception of a jackhammer sound, and the action associated with this source is restricted to "pounding." In contrast, other prompts, such as "a dog barking," benefit from more detailed descriptions, where variations in size (e.g., "small dog" vs. "large dog") or action (e.g., "barking" vs. "whining") could yield perceptually different audios. To maintain consistency across the dataset, we empirically balanced the complexity of foreground prompts, acknowledging that certain sounds carry more inherent information and, therefore, do not necessitate additional qualifiers or actions.

A sound engineer from our team created 4-s audio files corresponding to each prompt based on sounds sourced mainly from \url{Freesound.org} but also from private libraries. In total, our dataset comprises 310 audio-captions, with approximately 50 in each foreground sound category and 60 per background category. 
The development and evaluation set contain respectively 60 and 250 of these audio-caption pairs. Two background categories, "no background" and "birds," are excluded from the development set. Consequently, the evaluation set contains more samples with "no background" and "birds" compared to the development set.

\textbf{Baseline System} We provided AudioLDM \cite{audioldm} as our baseline model.
To ensure high quality and controllability, 9k hours of audio from 4 different sources were used for training. In addition, the model leveraged techniques such as the latent diffusion model and pretrained audio-text embedding \cite{clap_ms}, which made the training efficient.
As the baseline model generates 10-second audio, which is longer than our configuration, we 1) chopped audio into 4-second segments with a hop size of 2 seconds and selected the largest energy segment, and 2) resampled it from 16kHz to 32kHz.

\section{Evaluation}
\label{sec:evaluation}
Following the previous challenge edition
\cite{dcase2023}
, we conducted a two-stage evaluation scheme including both objective and subjective evaluation.

\textbf{Step 1: Objective metrics} To measure audio quality objectively, we adopted Fréchet Audio Distance (FAD) \cite{fad}. 
We chose FAD as it is a widely used metric in audio generation to measure set-wise audio quality and semantics compared to the reference set.
For the embedding used in FAD, we used PANNs CNN14 Wavegram-Logmel \cite{panns} (denoted as FAD-P) since it showed the highest correlation scores with perceptual rating \cite{fad_audio,fad_music}.
We provided an official evaluation software. 
\footnote{
\url{https://github.com/DCASE2024-Task7-Sound-Scene-Synthesis/fadtk}
}

\textbf{Step 2: Subjective metrics} Subjective evaluation of audio fit ("how well the audio matches the sound of the prompt") and perceptual quality ("clarity, absence of artifacts and distortion") was performed for the four submitted systems, the provided baseline system, and the Sound-Designer Reference evaluation set. Four prompts from each of the six foreground categories were selected, spanning the five background categories. First, 148 randomly ordered trials were presented online (via the toolkit \url{Gorilla.sc}) in six sections separated by foreground category. Category orders were varied across raters. 
Each audio was given a separate rating for its match to the foreground and background portion of the prompt on a scale from 0 (extremely poor) to 10 (extremely good). Subsequently, the same 148 sounds were presented in random order, without a prompt, and were rated for perceptual quality (0-10) regardless of content. Rating one sound per trial was better suited to this purpose than comparing multiple sounds because each sound was unique ~\cite{mushra}.

Fourteen raters, four from each top team and ten from system-blinded organizers and their lab members, rated sounds from all systems. To avoid bias, for each contestant and each prompt, each self-rating was replaced with a contestant’s average responses to that prompt for all other systems; replacement ensured that simple removal of self-ratings would not uniquely raise or lower a system’s average. The Final Rating of each system is a weighted sum of its Foreground Fit, Background Fit, and Audio Quality in a 2:1:1 ratio.

\section{Results}
\label{sec:results}

\begin{figure}[t!]
    \centering 
    \def\figwidth{0.25\textwidth}
    \begin{subfigure}{\figwidth}
        \centering
        \includegraphics[width=\linewidth]{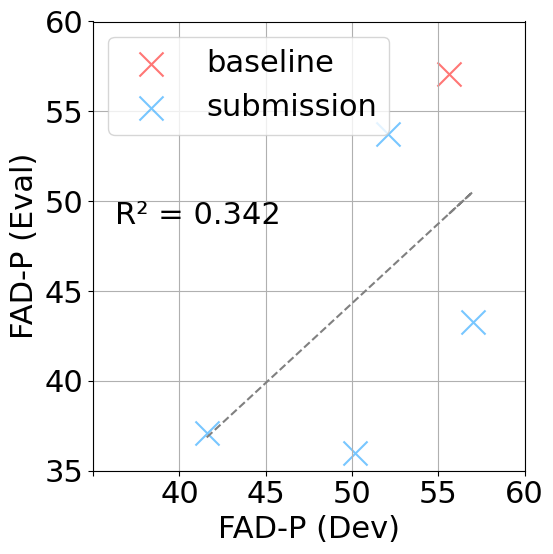}
        \subcaption{FAD-P on Evaluation set vs Development set}
        \label{fig:fad-dev-eval}
    \end{subfigure}
    \begin{subfigure}{\figwidth}
        \centering
        \includegraphics[width=\linewidth]{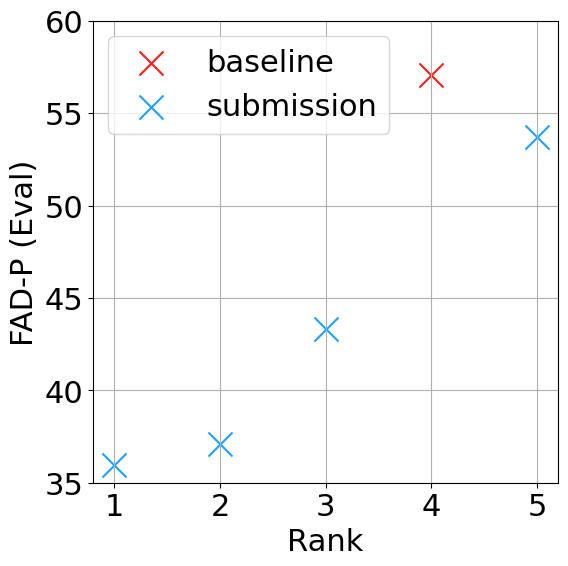}
        \subcaption{FAD-P on Evaluation set vs Challenge Ranking}
        \label{fig:fad-rank}
    \end{subfigure}
    \begin{subfigure}{0.34\textwidth}
        \centering
        \includegraphics[width=\linewidth]{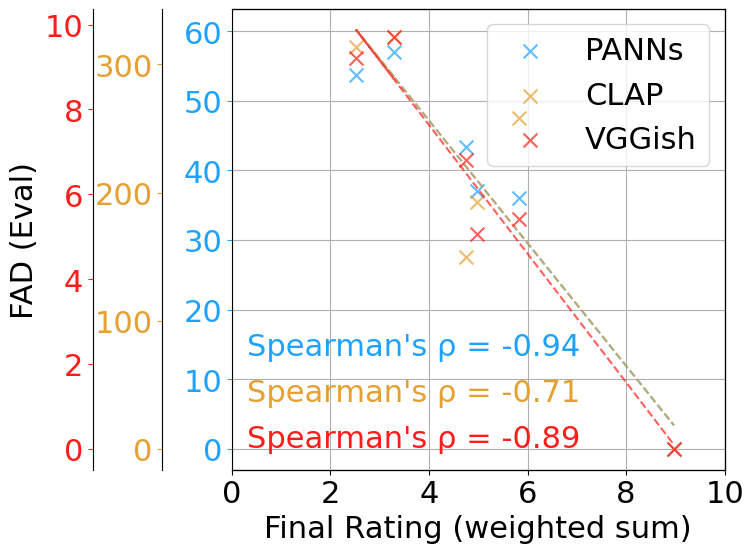}
        \subcaption{FAD scores from different embeddings vs Final Rating}
        \label{fig:fad_corr}
    \end{subfigure}
    \\
    \begin{subfigure}{\figwidth}
        \centering
        \includegraphics[width=\linewidth]{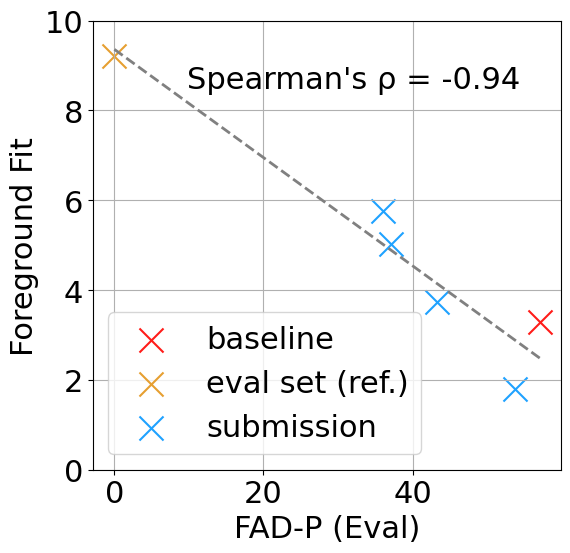}
        \subcaption{FAD-P on Evaluation set vs Foreground Fit}
        \label{fig:eval-mos-fg}
    \end{subfigure}
    \begin{subfigure}{\figwidth}
        \centering
        \includegraphics[width=\linewidth]{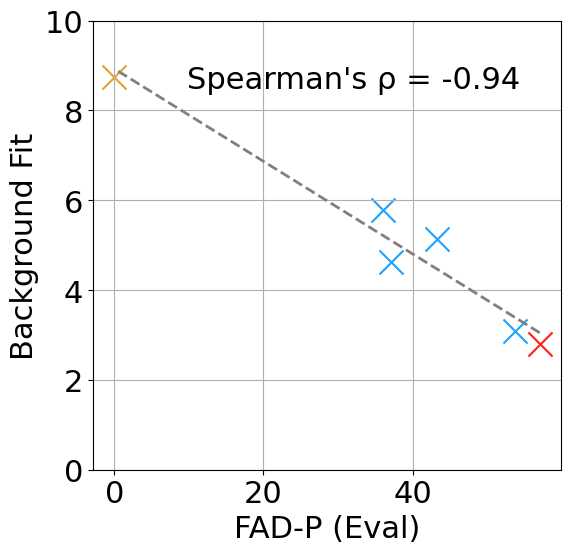}
        \subcaption{FAD-P on Evaluation set vs Background Fit}
        \label{fig:eval-mos-bg}
    \end{subfigure}
    \begin{subfigure}{\figwidth}
        \centering
        \includegraphics[width=\linewidth]{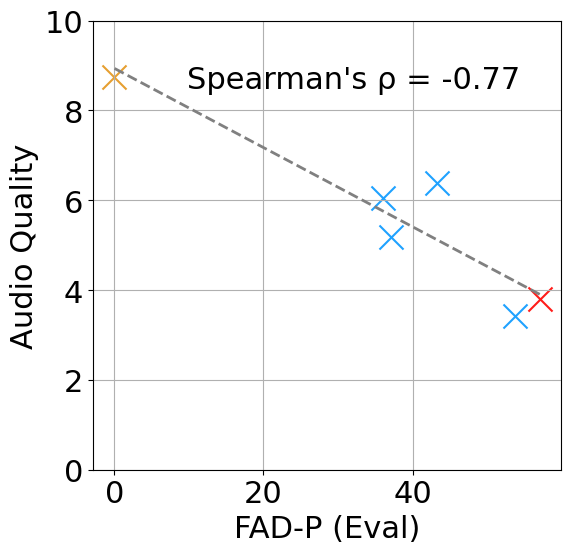}
        \subcaption{FAD-P on Evaluation set vs Audio Quality}
        \label{fig:eval-mos-aq}
    \end{subfigure}
    \caption{Correlation between FAD scores on evaluation set and other indicators, computed on the 4 submitted systems and the baseline system.}
    \label{}
\end{figure}

\begin{figure}[t!]
    \centering
    \resizebox{0.8\textwidth}{!}{%
        \includegraphics{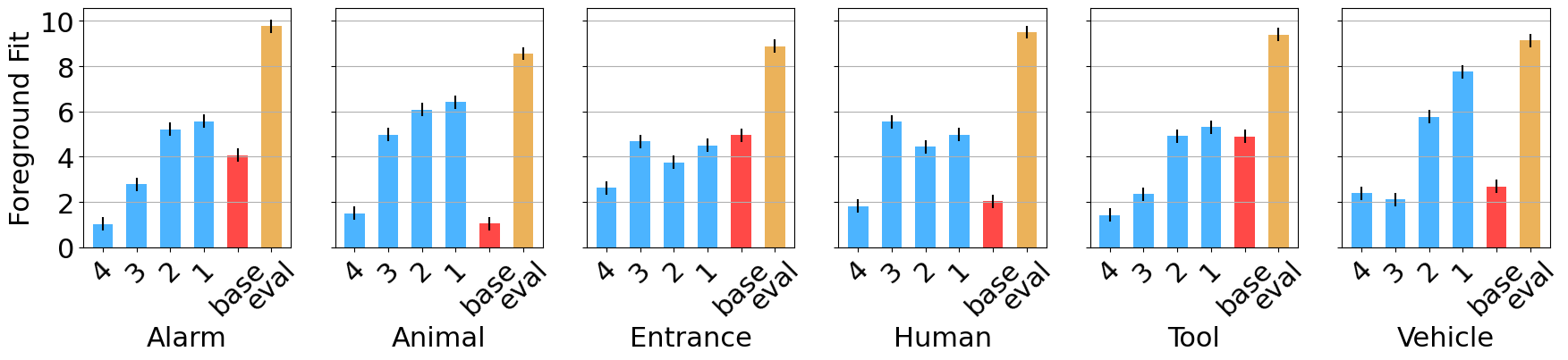}
    }
    \caption{Subjective evaluation results on Foreground Fit. The error bar indicates the standard error.}
    \label{fig:foreground}
\end{figure}

A total of four systems were received for submission \cite{Sun_Samsung,Chung_KT,Yi_Surrey,Verma_IITM}. 
In Figure \ref{fig:fad-dev-eval}, the FAD scores of 4 systems and the baseline system are plotted. The (x, y) position represents the FAD score computed on the development set (FAD-P Dev) and the evaluation set (FAD-P Eval), respectively. 
First, the majority of systems exhibit a tendency to achieve lower FAD-P scores on the evaluation set when they are lower on the development set, with the exception \cite{Yi_Surrey,Sun_Samsung}. This is anticipated, as the training is based, at least in part, on the development set. 
Second, it turns out that FAD-P Dev is a noisy measure to predict FAD-P Eval. 
The exception may be attributed to the presence of new sound sources in the evaluation set to prevent overfitting, which may result in performance discrepancies between the two sets.

Figure \ref{fig:fad-rank} shows the final rankings of the 4 systems and the baseline system, as determined by the weighted summed score from the listening test, in conjunction with the FAD-P Eval and FAD-P Dev. The FAD-P Eval scores align well with the final rankings, while FAD-P Dev does not. The Spearman's correlation coefficient $\rho$ of the ranking by FAD-P Eval and the final ranking is '0.900' ($p=0.037$), while by FAD-P Dev it is '0.500' ($p=0.391$). This discrepancy may be due to systems being overfitted to the development set or to the relatively small size of the development set since FAD is a biased metric \cite{fid_bias,fad_music}.

To validate the use of PANNs embedding in FAD calculation, we examined the correlation between FAD scores calculated from different embeddings and weighted summed scores, as illustrated in Figure \ref{fig:fad_corr}. The PANNs model demonstrated the highest Spearman's correlation coefficient of -0.94, in comparison to CLAP \cite{clap_ms} and VGGish \cite{vggish}. It is noteworthy that only the result of PANNs was statistically significant (i.e., $p<0.05$). This result corroborates the previous study's findings \cite{fad_audio}. 

Figure ~\ref{fig:eval-mos-fg} to ~\ref{fig:eval-mos-aq} illustrate the correlation between FAD-P and human subjective ratings. FAD-P shows a strong relationship with both foreground and background fit but a weak correlation with overall audio quality. This suggests that FAD-P primarily measures the audio-text correspondence, while it may be less sensitive to factors affecting overall quality, such as noise or generated artifacts.

To apply our dataset for evaluation, we additionally measured the FAD-P on the evaluation set with generated results from other open-source models \cite{audioldm2,make-an-audio,consistencytta,tango,tango2,tango-llm,auffusion,stableaudio,audiogen} (see Figure \ref{fig:fad-models}). 
Our findings revealed a consistent trend whereby scaling up resulted in enhanced performance, which aligns with the prevalent notion in the field of generative models. The number of model parameters was a more dominant factor than the model types (transformers or diffusion models) in general. However, there is one notable exception: Chung\_KT \cite{Chung_KT} demonstrated promising performance in a lightweight GAN-based architecture. Secondly, it was observed that a model generating audio at a higher sample rate did not always achieve a better score in the evaluation set at 32 kHz. Currently, it is relatively under-optimized to train a model with a higher and more production-ready sampling rate. 

The mean subjective ratings of Foreground fit, Background fit, and Audio Quality were appropriately low for the baseline system (3.3, 2.8, 3.8), appropriately high for the Reference Set (9.8, 8.8, 9.0), and moderately high for the top-ranked submitted system (5.8, 5.8, 6.0) \footnote{\url{https://dcase.community/challenge2024/task-sound-scene-synthesis-results}
}. Figure \ref{fig:foreground} shows the mean Foreground Fit ratings (and their standard errors, calculated over the distribution of 14 ratings, showing high interrater agreement, Cronbach’s $\alpha$ = 0.959) for each submission within each foreground category. 
The Background  Fit (not shown) correlates highly (r=0.79) with the Foreground Fit, and Audio Quality correlates highly with  both Foreground Fit (r=0.85) and Background Fit (r=0.87). Although system rankings vary across categories, and different rankings do not always reflect a large mean difference, the overall winner (submission 1) has the highest rating in most of the foreground categories. The Entrance Category proved the most challenging for generative systems, with no submissions rated a higher fit than the baseline system, while all submissions fared better than the baseline system in the Animal Category.

\section{Conclusion}
\label{sec:conclusion}
The Sound Scene Synthesis challenge has yielded important insights into text-to-audio generation for environmental sounds. Our evaluation protocol, combining FAD-P metrics and human ratings, revealed both progress and areas for improvement in audio quality, diversity, and controllability. The structured prompt format facilitated diverse captions while enabling effective evaluation. While larger models generally excelled, innovative lightweight approaches also showed promise. Performance varied across sound categories, with some showing substantial improvement over the baseline. The strong correlation between FAD-P and human ratings, particularly for sound source fit, validates its use as a reliable objective metric for future research.

Future work should focus on enhancing the nuance, temporal aspects, and spatial capabilities of generated sounds. Refining evaluation metrics to capture subtle qualitative differences will be crucial. As this task serves as a valuable benchmark for assessing generative audio models, future iterations could incorporate more sophisticated prompts and criteria. Success in this domain could pave the way for more complex audio generation tasks such as video-to-audio synthesis, potentially revolutionizing AI-driven audio production for multimedia content.


{
\small
\bibliography{ref}
\bibliographystyle{unsrt} 

}

\newpage
\appendix

\section{Challenge Task Overview}

\begin{figure}[h!]
    \centering
    \resizebox{0.8\textwidth}{!}{%
        \includegraphics{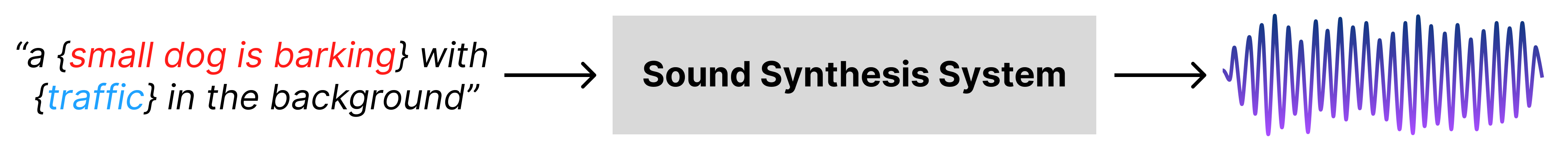}
    }
    \caption{Overview of \textit{Sound Scene Synthesis} task. A sound synthesis system (i.e., Text-to-Audio model) receives a text prompt as an input, and outputs an audio corresponding to the prompt.}
    \label{fig:task}
\end{figure}

\begin{figure}[h!]
    \centering
    \resizebox{0.38\textwidth}{!}{%
        \includegraphics{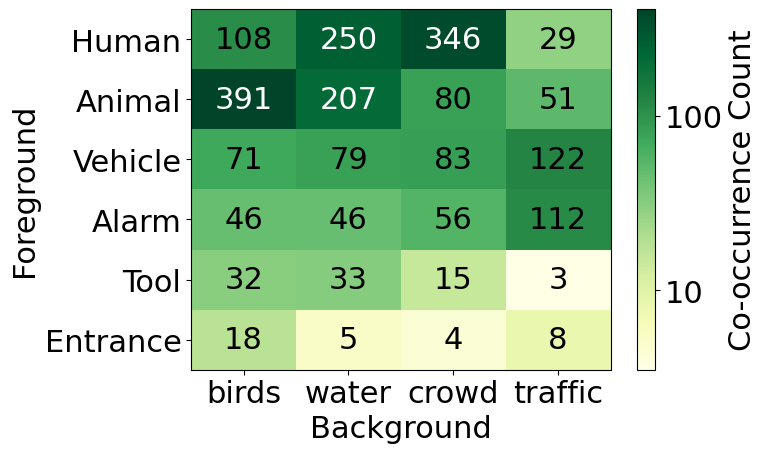}
    }
    \caption{Heatmap of the co-occurrence of foreground-background combinations in AudioCaps\cite{audiocaps} trainset. The data imbalance may potentially limit the model's controllability.}
    \label{fig:co-occurrence}
\end{figure}

\section{Challenge Results}

\begin{figure}[h!]
    \centering
    \resizebox{0.8\textwidth}{!}{%
        \includegraphics{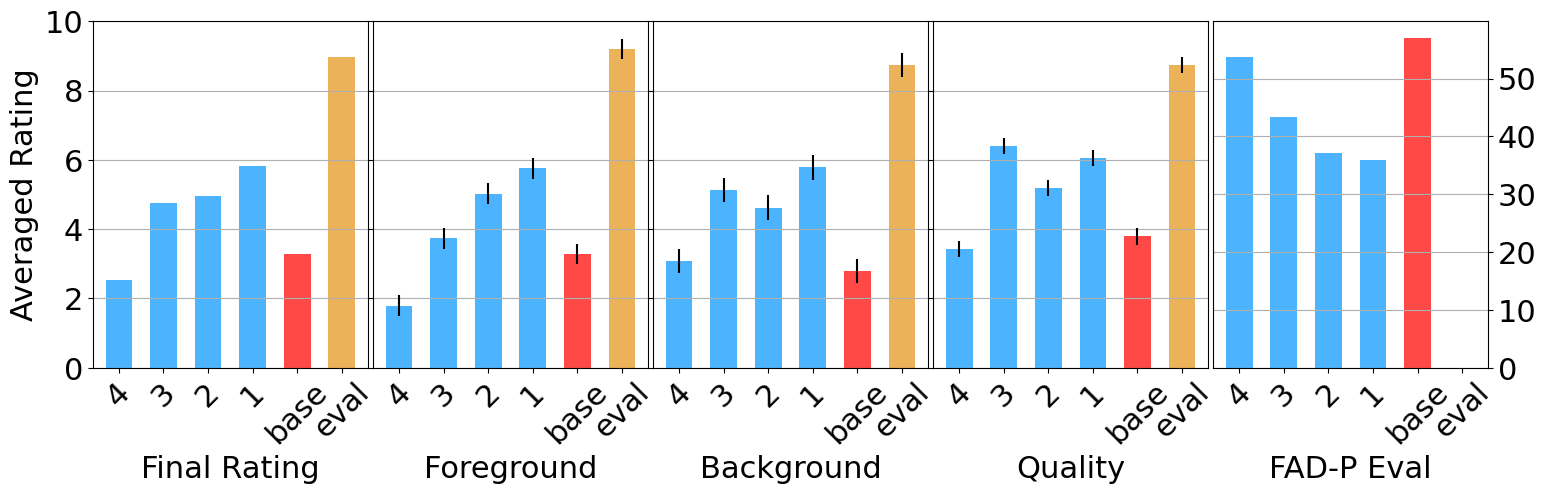}
    }
    \caption{    Evaluation score for each system averaged across all sounds: Far left panel, Final Rating, combines subjective ratings of Foreground: Background: Quality with 2:1:1 weighting. Far right panel, Objective evaluation score (FAD-P Eval). The error bar indicates the standard error.
    The official ranking is as follows: $1^{st}$ \textit{Sun\_Samsung}\cite{Sun_Samsung}, $2^{nd}$ \textit{Chung\_KT}\cite{Chung_KT}, $3^{rd}$ \textit{Yi\_Surrey}\cite{Yi_Surrey}, $4^{th}$ \textit{Verma\_IITM}\cite{Verma_IITM}.}
    \label{fig:average}
\end{figure}

\begin{figure}[h!]
    \centering
    \resizebox{0.8\textwidth}{!}{%
        \includegraphics{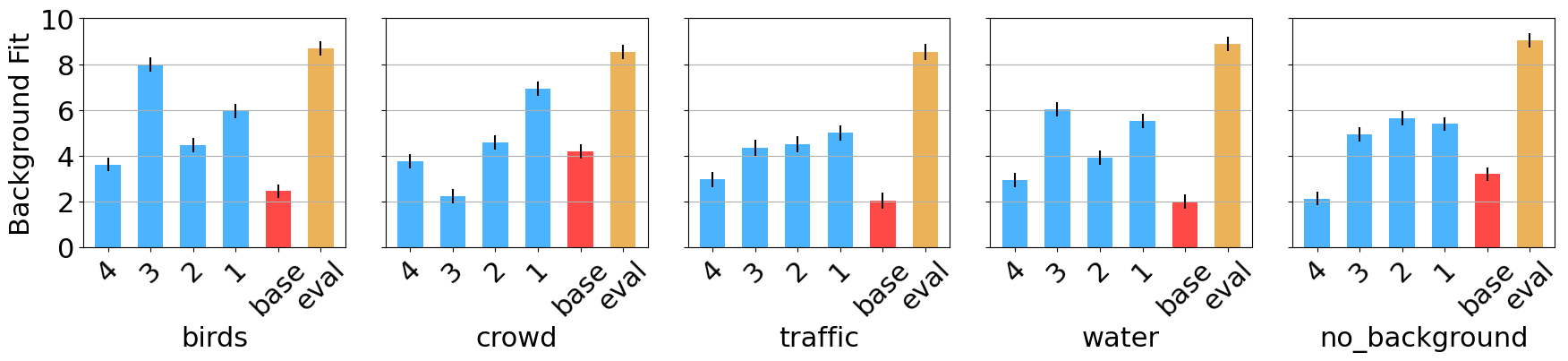}
    }
    \caption{Subjective evaluation results on Background Fit. The error bar indicates the standard error.}
    \label{fig:background}
\end{figure}

\begin{figure}[h!]
    \centering
    \resizebox{0.8\textwidth}{!}{%
        \includegraphics{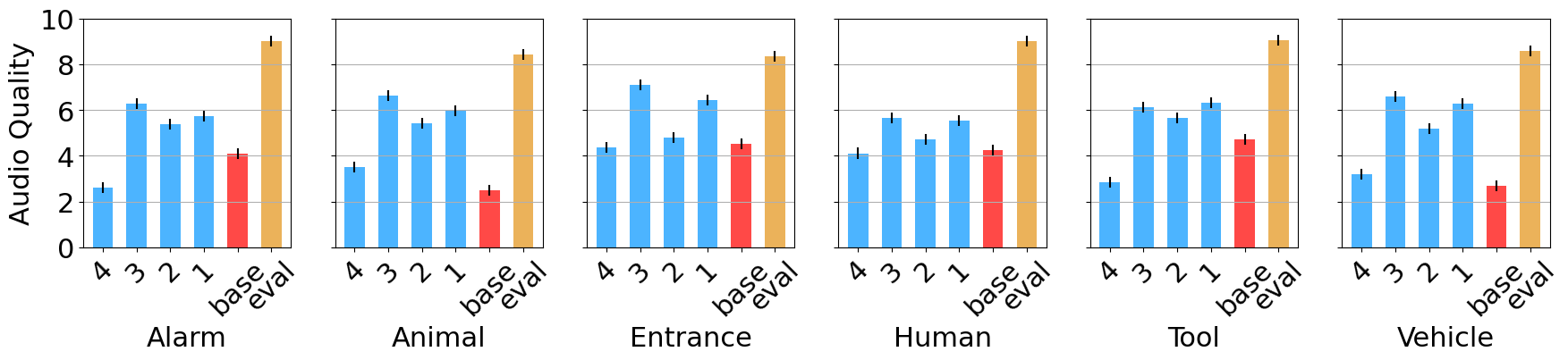}
    }
    \caption{Subjective evaluation results on Audio Quality. The error bar indicates the standard error.}
    \label{fig:quality}
\end{figure}

These additional figures are provided to display the challenge results on subjective evaluation. 
The x-axis indicates the official ranking, where "base" refers to the baseline system and "eval" denotes the reference evaluation set created by a sound designer.
Figure \ref{fig:average} depicts the Final Rating (i.e., weighted sum as outlined in Section \ref{sec:evaluation}), in conjunction with other averaged scores and FAD-P. Note that unlike other metrics, a lower FAD-P score means better performance and the FAD-P for "eval" is zero.
Figure \ref{fig:background} illustrates the mean Background Fit within each background category.
Figure \ref{fig:quality} shows the mean Audio Quality within each foreground category.
The resulting inter-rater agreement, among 14 raters over 144 prompts, was high (Cronbach’s $\alpha=$ 0.959).
Please refer to our challenge homepage for further detailed results and numerical data. \footnote{\url{https://dcase.community/challenge2024/task-sound-scene-synthesis-results}}

\section{Subjective Evaluation}
\label{apdx:subjective_eval}

\begin{figure}[h!]
    \centering
    \def\figwidth{0.48\textwidth}
    \begin{subfigure}{\figwidth}
        \centering
        \includegraphics[width=\linewidth]{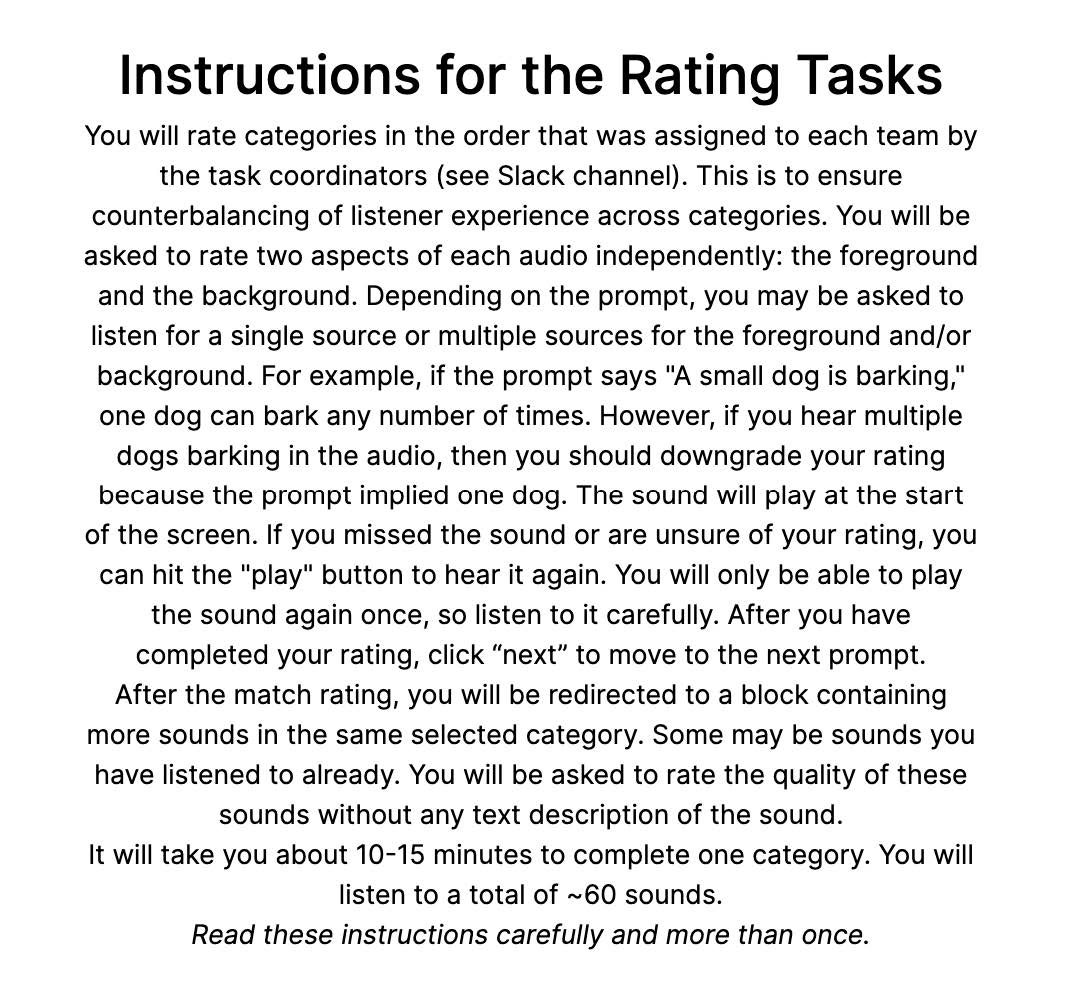}
    \end{subfigure}
    \\
    \begin{subfigure}{\figwidth}
        \centering
        \includegraphics[width=\linewidth]{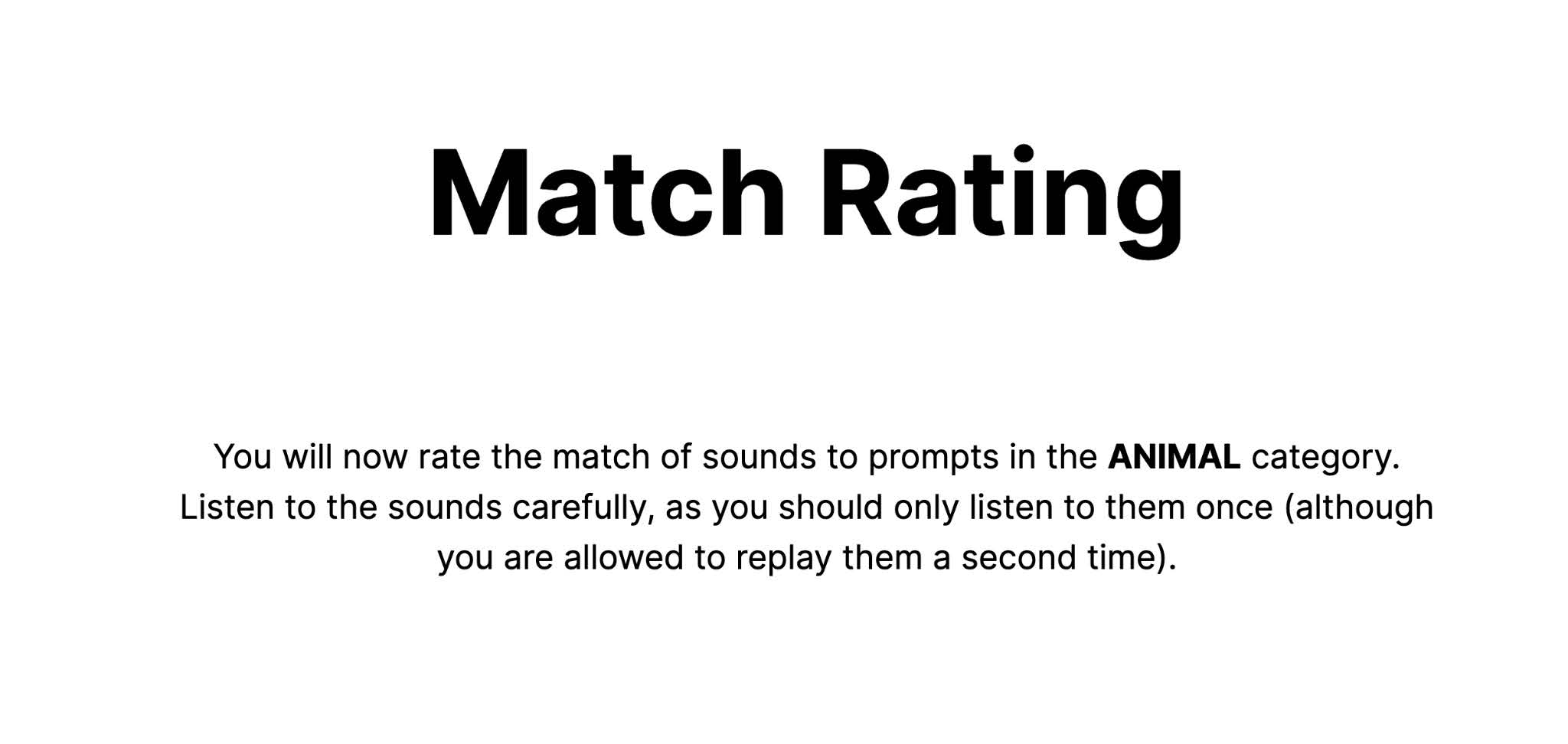}
    \end{subfigure}
    \begin{subfigure}{\figwidth}
        \centering
        \includegraphics[width=\linewidth]{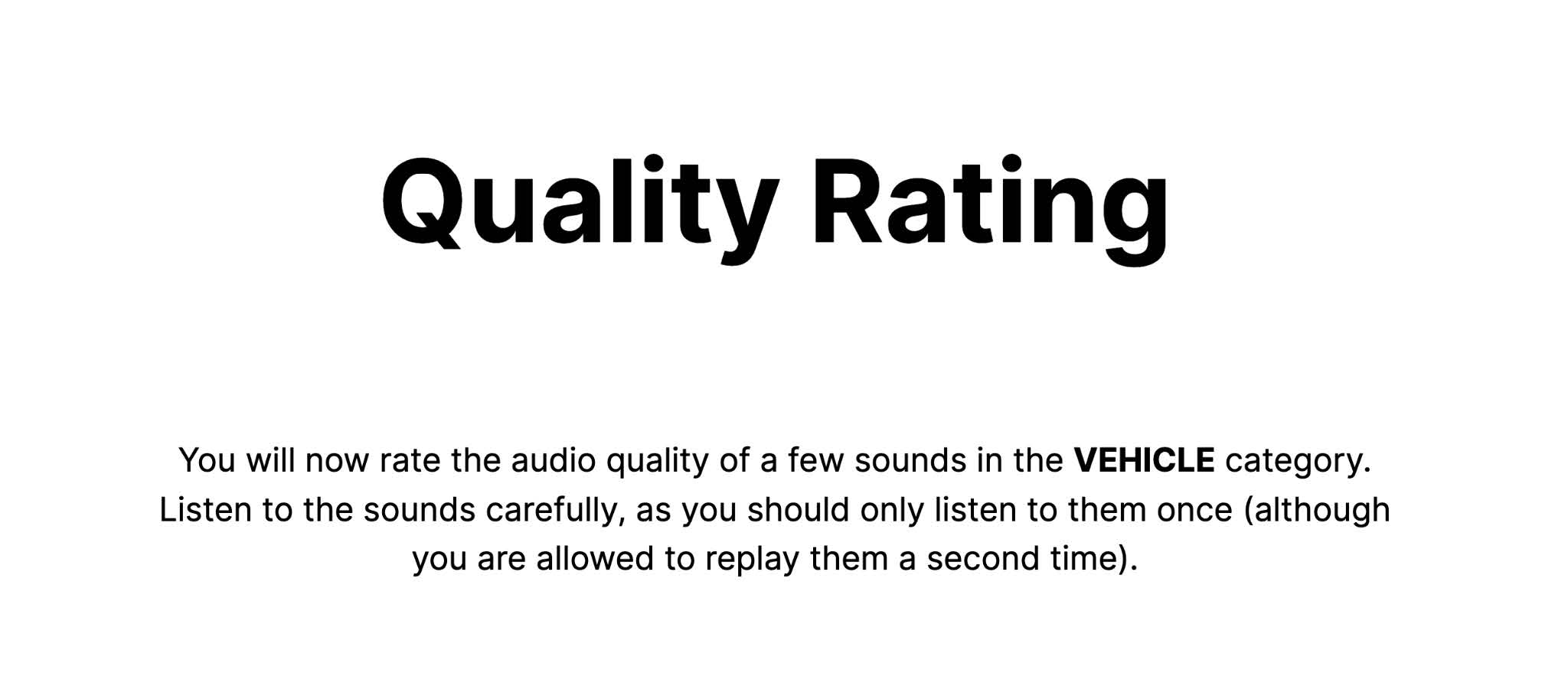}
    \end{subfigure}
    \\
    \begin{subfigure}{\figwidth}
        \centering
        \includegraphics[width=\linewidth]{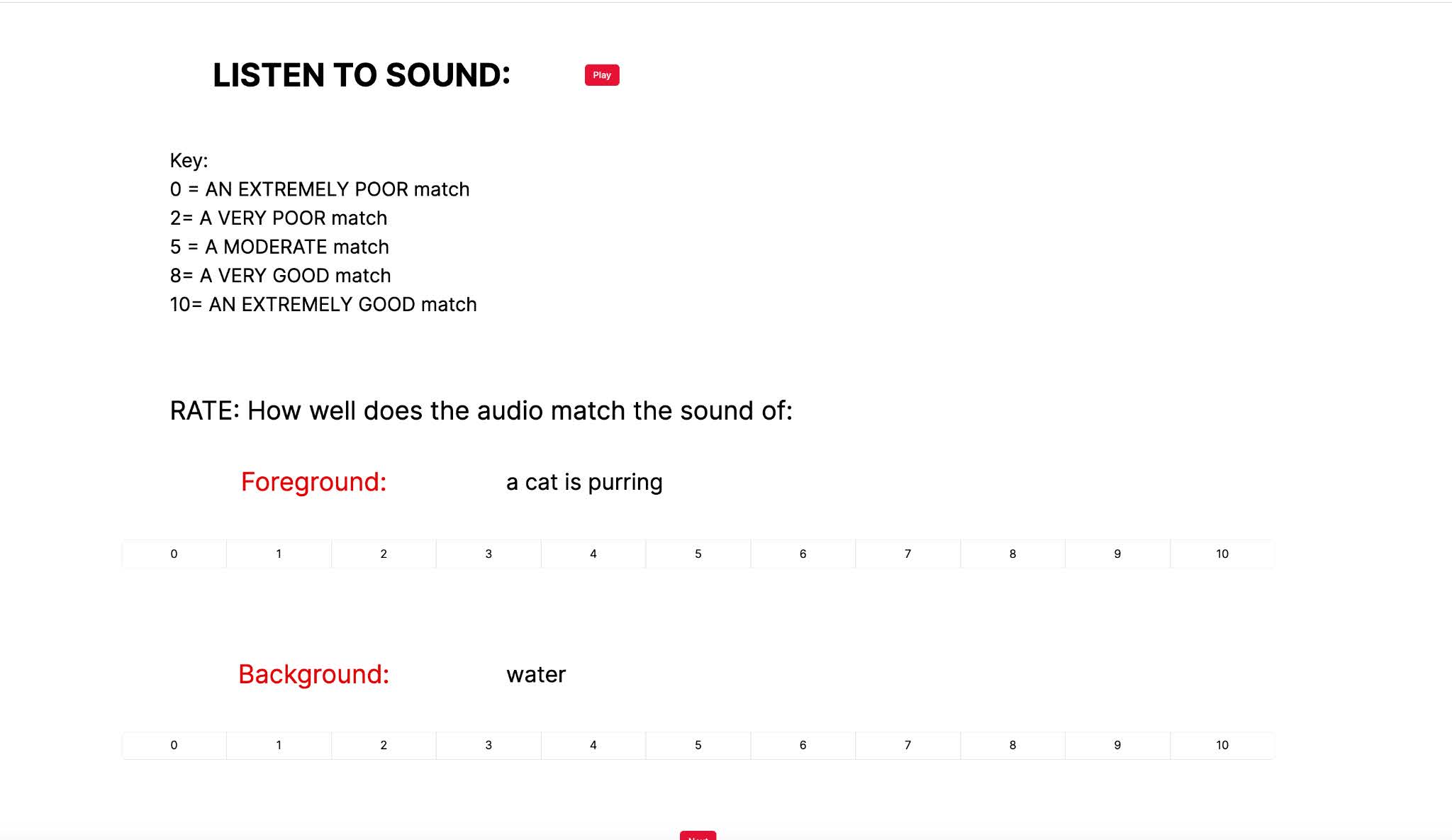}
    \end{subfigure}
    \begin{subfigure}{\figwidth}
        \centering
        \includegraphics[width=\linewidth]{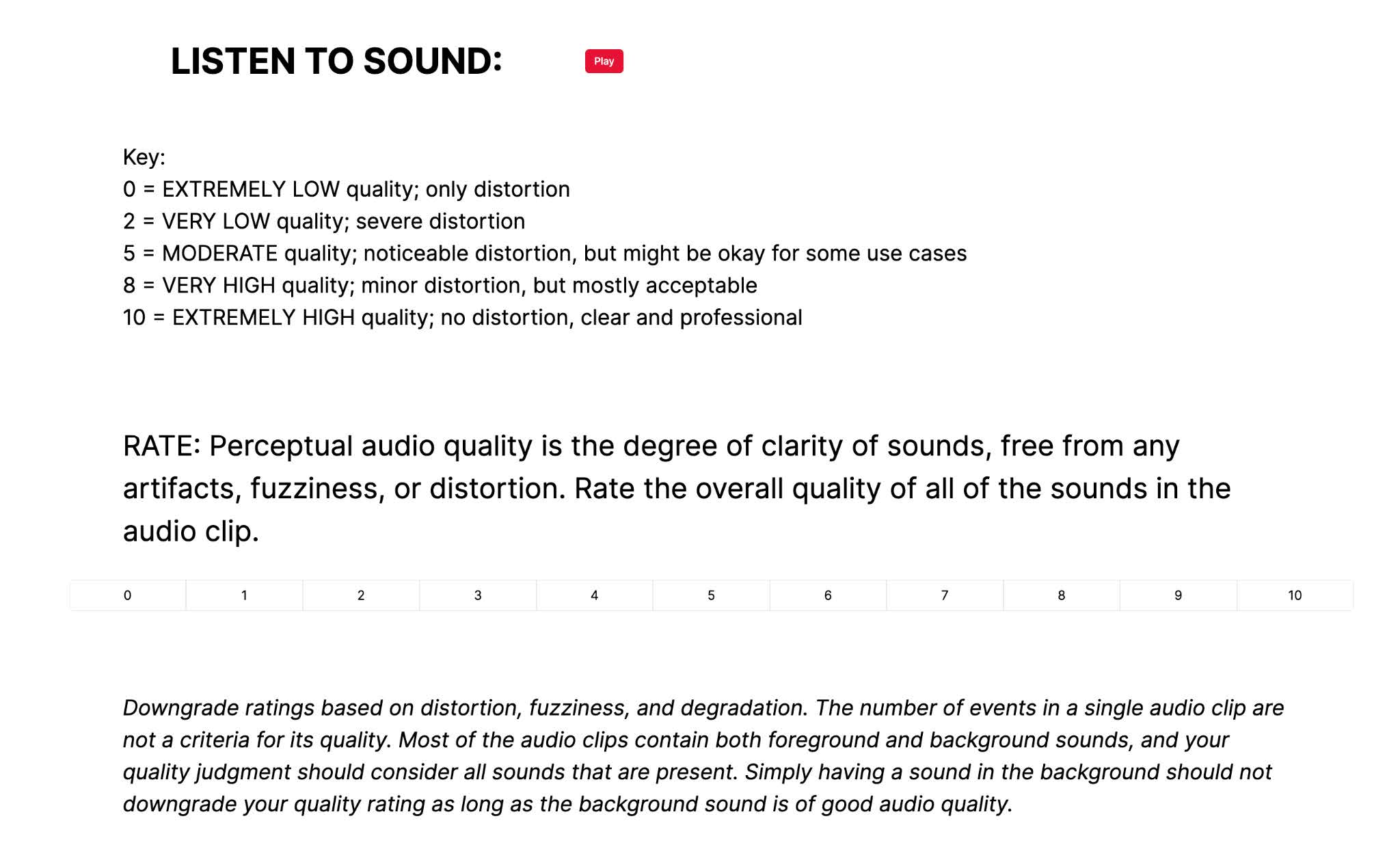}
    \end{subfigure}
    \caption{Screenshots of subjective evaluation toolkit}
    \label{fig:subjective_eval}
\end{figure}

Figure \ref{fig:subjective_eval} shows the screenshot of the platform used for subjective evaluation. See Section \ref{sec:evaluation} for more details.



\newpage

\end{document}